\documentclass[aps,prd,superscriptaddress,showpacs,preprint]{revtex4}
\usepackage{graphicx, bm}
\begin{document}
\draft
\title{Bounding the Number of Light Neutrinos Species in a Left-Right Symmetric Model}

\author{ A. Guti\'errez-Rodr\'{\i}guez}
\affiliation{\small Facultad de F\'{\i}sica, Universidad Aut\'onoma de Zacatecas\\
         Apartado Postal C-580, 98060 Zacatecas, Zacatecas M\'exico.\\
         Cuerpo Acad\'emico de Part\'{\i}culas Campos y Astrof\'{\i}sica.}
\author{M. A. Hern\'andez-Ru\'{\i}z}
\affiliation{\small Facultad de Ciencias Qu\'{\i}micas, Universidad Aut\'onoma de Zacatecas\\
        Apartado Postal 585, 98060 Zacatecas, Zacatecas M\'exico.\\}
\author{M. A. P\'erez}
\affiliation{\small Departamento de F\'{\i}sica, CINVESTAV.\\
             Apartado Postal 14-740, 07000, M\'exico D.F., M\'exico.}
\author{F. P\'erez-Vargas $^1$}

\date{\today}

\begin{abstract}

Using the experimental values for the rates $R^{LEP}
_{exp}=\Gamma_{inv}/\Gamma_{l\bar l}=5.942\pm 0.016$,
$R^{Giga-Z_1}=\Gamma_{inv}/\Gamma_{l\bar l}=5.942\pm 0.012$ (most
conservative) and $R^{Giga-Z_1}=\Gamma_{inv}/\Gamma_{l\bar
l}=5.942\pm 0.006$ (most optimistic) we derive constraints on the
number of neutrino light species $(N_\nu)_{LRSM}$ with the
invisible width method in the framework of a left-right symmetric
model (LRSM) as a function of the LR mixing angle $\phi$. Using
the LEP result for $N_\nu$ we may place a bound on this angle,
$-1.6\times 10^{-3}\leq\phi \leq 1.1\times 10^{-3}$, which is
stronger than those obtained in previous studies of the LRSM.
\end{abstract}

\pacs{14.60.Lm,12.15.Mm, 12.60.-i
\vspace*{2cm}\noindent  E-mail: $^{1}$alexgu@planck.reduaz.mx,
$^{2}$mahernan@uaz.edu.mx, $^{3}$mperez@fis.cinvestav.mx}

\vspace{5mm}

\maketitle


\section{Introduction}

The number of fermion generations, which is associated to the
number of light neutrinos, is one of the most important
predictions of the Standard Model of the electroweak interactions
(SM) \cite{S.L.Glashow}. In the SM the decay width of the $Z_1$
boson into each neutrino family is calculated to be
$\Gamma_{\nu\bar \nu}= 166.3\pm 1.5$ $MeV$ \cite{Data06}.
Additional generations, or other new weakly interacting particles
with masses below $M_{Z_1}/2$, would lead to a decay width of the
$Z_1$ into invisible channels larger than the SM prediction for
three families while a smaller value could be produced, for
example, by the presence of one or more right-handed neutrinos
mixed with the left-handed ones \cite{Jarlskog}. Thus the number
of light neutrino generations $N_\nu$, defined as the ratio
between the measured invisible decay width of the $Z_1$,
$\Gamma_{inv}$, and the SM expectation $\Gamma_{\nu\bar \nu}$ for
each neutrino family, need not be an integer number and has to be
measured with the highest possible accuracy.

The most precise measurement of the number of light $(m_\nu < 45
\hspace{1mm} GeV)$ active neutrino types, and therefore the number
of associated fermion families, comes from the invisible $Z_1$
width $\Gamma_{inv}$, obtained by subtracting the observed width
into quarks and charged leptons from the total width obtained from
the lineshape. The number of effective neutrinos $N_\nu$ is given
by \cite{M.Carena}

\[
N_\nu =\frac{\Gamma_{inv}}{\Gamma_l}(\frac{\Gamma_{l\bar
l}}{\Gamma_\nu})_{SM},
\]

\noindent where $(\frac{\Gamma_{l\bar l}}{\Gamma_\nu})_{SM}$, the
SM expression for the ratio of widths into a charged lepton and a
single active neutrino, is introduced to reduce the model
dependence. The experimental value from the four LEP experiments
is $N_\nu$= 2.9841$ \pm $0.0083 \cite{Data06,Abbaneo}, excluding
the possibility of a fourth family unless the neutrino is very
heavy or sterile. $N_\nu$ is the effective number of light
neutrino generations deduced from the $Z_1$ invisible width based
on the expected partial width for one light neutrino generation
($N_\nu=\Gamma_{inv}/\Gamma^{SM}_\nu$). This result is in
agreement with cosmological constraints on the number of
relativistic species around the time of Big Bang nucleosynthesis,
which seems to indicate the existence of three very light neutrino
species \cite{V.Barger}. On the other hand, the LEP result
measures precisely the slight deviations of $N_\nu$ from three. In
particular, the most precise LEP numbers can be translated into
$N_\nu$ = 2.9841$\pm 0.0083$ \cite{Data06}, about two sigma away
from the SM expectation, $N_\nu$= 3. While not statistically
significant, this result suggests that the $Z\nu\bar\nu$-couplings
might be suppressed with respect to the SM value
\cite{M.Carena,M.Maya}.

Using the experimental value for
$R^{LEP}_{exp}=\frac{\Gamma_{inv}}{\Gamma_{l\bar l}}=5.942\pm
0.016$ \cite{Abbaneo}, we will determine the allowed region for
$(N_\nu)_{LRSM}$ as a function of the mixing angle $\phi$ and
estimate bounds for the number of light neutrinos species in the
framework of a left-right symmetric model (LRSM)
\cite{A.Gutierrez,J.C.Pati}. We will also use the LEP results to
get a constraint on the LR  mixing angle $\phi$. On the other
hand, if one assumes that the results for $\Gamma_{inv}$ and
$\Gamma_{l\bar l}$ for a future TESLA-like Giga-$Z_1$ experiment
agree with the central values obtained at LEP, one would measure
$(\frac{\Gamma_{inv}}{\Gamma_{l\bar l}})^{Giga-Z_1}=5.942\pm
0.012$ (most conservative) or $(\frac{\Gamma_{inv}}{\Gamma_{l\bar
l}})^{Giga-Z_1}=5.942\pm 0.006$ (most optimistic) \cite{M.Carena},
in this case we estimate also a limit for the number of light
neutrinos species.

This paper is organized as follows: In Sec. II we present the
expressions for the decay widths of $Z_1 \to l\bar l$ and $Z_1 \to
\nu \bar \nu$ in the LRSM. In Sec. III we present the numerical
computation and, finally, we summarize our results in Sec. IV.

\section{Widths of $Z_1 \to l\bar l$ and $Z_1 \to \nu \bar \nu$}

In this section we calculate the partial widths for $ Z_1 \to
l\bar l$ and $ Z_1 \to \nu \bar \nu$ using the transition
amplitude given in Ref. \cite{A.Gutierrez} in the context of the
LRSM. The expression for the transition amplitude for the channel
$Z_1 \to l\bar l$ is given by

\begin{equation}
M(Z_1 \rightarrow l\bar l)= \frac{g}{cos\theta_W}[\bar
u(l)\gamma^u\frac{1}{2}(ag_V^l - bg_A^l\gamma_5) v(\bar
l)]\varepsilon_\mu^\lambda(Z_1),
\end{equation}

\noindent where $u(v)$ is the lepton (antilepton) spinor and
$\varepsilon_\mu^\lambda$ is the $Z_1$ boson polarization vector
and the expressions for the couplings $a$ and $b$ in the LRSM are:

\begin{equation}
a=\cos{\phi}-\frac{\sin{\phi}}{\sqrt{\cos 2\theta_W}} \hspace{5mm}
\mbox{and} \hspace{5mm} b=\cos{\phi}+\sqrt{ \cos2\theta_
W}\sin{\phi},
\end{equation}

\noindent where $\phi$ is the mixing angle of the LRSM
\cite{M.Maya,J.Polak}.

After applying some of the trace teorems of the Dirac matrices and
of sum and average over the initial and final spin the square of
the matrix elements becomes

\begin{equation}
\Sigma_s |M|^2 =
\frac{g^2M_{Z_1}^2}{3cos^2\theta_W}[a^2(g_V^l)^2(1+\frac{2m_l^2}
{M^2_{Z_1}}) + b^2(g_A^l)^2(1-\frac{4m_l^2}{M^2_{Z_1}})].
\end{equation}

Our next step, now that we know the square of the Eq. (3)
transition amplitude, is to calculate the partial width of the
reaction $Z_1 \to l\bar l$:

\begin{eqnarray}
\Gamma_{l\bar l} =
\frac{G_FM_{Z_1}^3}{6\pi\sqrt{2}}\sqrt{1-4\eta_l}[a^2(g_V^l)^2 +
b^2(g_A^l)^2 + 2\eta_l(a^2(g_V^l)^2- 2b^2(g_A^l)^2)],
\end{eqnarray}

\noindent where  $\eta_l=\frac{m^2_l}{M^2_{Z_1}}$.

For the $Z_1$-decay width into $\nu \bar \nu$ we obtain

\begin{eqnarray}
\Gamma_{\nu \bar \nu} =
\frac{G_FM^3_{Z_1}}{12\pi\sqrt{2}}\sqrt{1-4\eta_\nu}[\frac{1}{2}(a^2
+ b^2) + \eta_\nu (a^2-2b^2)],
\end{eqnarray}

\noindent where $\eta_\nu=\frac{m^2_\nu}{M^2_{Z_1}}$.

The partial widths Eqs. (4) and (5) are applicable to all charged
leptons and all neutrinos respectively.

\section{Results}

In order to compare the respective expressions Eqs. (4) and (5)
with the experimental result for the number of light neutrinos
species $N_\nu$, we will use the definition for $N_\nu$ in a SM
analysis \cite{M.Acciarri},

\begin{equation}
N_{\nu}=R_{exp}(\frac{\Gamma_{l\bar l}}{\Gamma_{\nu
\bar\nu}})_{_{SM}},
\end{equation}

\noindent where the quantity in parenthesis is the standard model prediction
and the $R_{exp}$ factor is the experimental value of the ratio
between the widths $\Gamma_{inv}$ and $\Gamma_{l\bar l}$
\cite{Data06,Abbaneo},

\begin{equation}
R^{LEP}_{exp}=(\frac{\Gamma_{inv}}{\Gamma_{l\bar l}})=5.942\pm
0.016.
\end{equation}

This definition replaces the expression
$N_\nu=\frac{\Gamma_{inv}}{\Gamma_{\nu \bar \nu}}$ since (7)
reduces the influence of the top quarks mass. To get information
about what is the meaning of $N_\nu$ in the LRSM we should define
the corresponding expression \cite{M.Maya},

\begin{equation}
(N_{\nu})_{LRSM}=R_{exp}(\frac{\Gamma_{l\bar l}}{\Gamma_{\nu
\bar\nu}})_{_{LRSM}}.
\end{equation}

This new expression is a function of the mixing angle $\phi$, so
in this case the quantity defined as the number of light neutrinos
species is not a constant and not necessarily an integer. Also,
$(N_\nu)_{LRSM}$ in formula (8) is independent from the $Z_2$ mass
and therefore depends only of the mixing angle $\phi$ of the LRSM.
Experimental values for $\Gamma_{inv}$ and for $\Gamma_{l\bar l}$
are reported in literature which, in our case, can give a bound
for the angle $\phi$. However, we can look to those experimental
numbers in another way. The partial widths $\Gamma_{inv}=499.0\pm
1.5$ $MeV$ and $\Gamma_{l\bar l}=83.984\pm 0.086$ $MeV$ were
reported recently \cite{Data06}, but we use the value given by (8)
for the $R_{exp}$ rate of Ref. \cite{Abbaneo}. All these
measurements are independent of any model and can be fitted with
the LRSM parameter $(N_\nu)_{LRSM}$ in terms of $\phi$.

In order to estimate a limit for the number of light neutrinos
species $(N_\nu)_{_{LRSM}}$ in the framework of a left-right
symmetric model, we plot the expression (8) to see the general
behavior of the $(N_\nu)_{_{LRSM}}$ function, Fig. 1. For the
mixing angle $\phi$ between $Z_1$ and $Z_2$, we use the reported
data of Maya {\it et al.} \cite{M.Maya}:

\begin{equation}
-9\times 10^{-3}\leq\phi \leq 4\times 10^{-3},
\end{equation}

\noindent with a $90\%$ C.L. Other limits on the mixing angle
$\phi$ reported in the literature are given in Refs.
\cite{J.Polak,Adriani}. In this figure we observed that for the
mixing angle $\phi$, around 0.65 rad, $(N_\nu)_{_{LRSM}}$ can be
as high as 5.9, and for values of $\phi$ around -0.95 rad,
$(N_\nu)_{_{LRSM}}$ is as low as 0. This shows a strong dependence
in $\phi$ for leptonic decays of the $Z_1$ boson. Therefore,
according to the above discussion, if we consider
$(N_\nu)_{_{LRSM}}$ as the number of neutrinos, the restriction on
the number of species can be ``softened" if we consider a LRSM. In
Fig. 2, we show the allowed region for $(N_\nu)_{_{LRSM}}$ as a
function of $\phi$ with $90\%$ C.L. The allowed region is the
inclined band that is a result of both factors in Eq. (8). In this
figure $(\frac{\Gamma_{l\bar l}}{\Gamma_{\nu \bar\nu}})_{_{LRSM}}$
gives the inclination while $R_{exp}$ gives the broading. This
analysis was done using the experimental value given in Eq. (7)
for $R_{exp}$ reported by \cite{Abbaneo} with a $90\%$ C.L. In the
same figure we show the SM $(\phi=0)$ result at $90\%$ C.L. with
the dashed horizontal lines.  The allowed region in the LRSM
(dotted line) for $(N_\nu)_{_{LRSM}}$ is wider that the one for
the SM, and is given by:

\begin{equation}
2.925\leq(N_\nu)_{_{LRSM}}\leq 3.02 \hspace{3mm} \mbox{or}
\hspace{3mm} (N_\nu)_{_{LRSM}}=2.987^{+0.033}_{-0.062},
\hspace{3mm} 90\% \hspace{1mm}C.L.,
\end{equation}

\noindent whose center value is quite close to the standard model
of three active neutrino species.

In the case of a future TESLA-like Giga-$Z_1$ experiment we obtain
the limits

\begin{eqnarray}
\hspace*{-5mm}2.926&\leq &(N_\nu)_{_{LRSM}}\leq 3.019 \hspace{2mm}
\mbox{or} \hspace{2mm} (N_\nu)_{_{LRSM}}=2.987^{+0.032}_{-0.061},
\hspace{2mm} 90\% \hspace{1mm}C.L.
\hspace{2mm} \mbox{(most conservative)}\\
\hspace*{-5mm}2.929&\leq &(N_\nu)_{_{LRSM}}\leq 3.016 \hspace{2mm}
\mbox{or} \hspace{2mm} (N_\nu)_{_{LRSM}}=2.987^{+0.029}_{-0.058},
\hspace{2mm} 90\% \hspace{1mm}C.L. \hspace{2mm} \mbox{(most
optimistic)},
\end{eqnarray}

\noindent which are consistent with those reported in the literature
\cite{M.Carena}.

Finally, and just for completeness, we reverse the arguments that
is, we fix the number of neutrinos in the LRSM to be three
then the theoretical expression for $R$ will be given by

\begin{equation}
R_{LRSM}=\frac{3\Gamma_{\nu\bar\nu}}{\Gamma_{l\bar l}}.
\end{equation}

The plot of this quantity as function of the mixing angle $\phi$
is shown in Fig. 3. The horizontal lines give the experimental
region at $90 \%$ C.L. From the figure we observed that the
constraint for the $\phi$ angle is:

\begin{equation}
-1.6\times 10^{-3}\leq\phi \leq 1.1\times 10^{-3},
\end{equation}

\noindent which is about one order of magnitude stronger than the
one obtained in previous studies of the LRSM \cite{M.Maya,J.Polak,Adriani}.

In the case of a future TESLA-like Giga-$Z_1$ experiment we would obtain
the following bounds for the mixing angle $\phi$:

\begin{eqnarray}
-1.1\times 10^{-3}&\leq &\phi\leq 0.9 \times 10^{-3}, \hspace{2mm} \mbox{(most conservative)},\\
-0.8\times 10^{-3}&\leq &\phi\leq 0.33 \times 10^{-3},
\hspace{2mm} \mbox{(most optimistic)}.
\end{eqnarray}

\section{Conclusions}

We have determined a bound on the number of light neutrinos
species in the framework of a left-right symmetric model as a
function of the mixing angle $\phi$, as shown in Eq. (10) and Fig.
2. Using this result and the LEP values obtained for $N_\nu$, we
were able to put a limit on the LR mixing angle $\phi$ which is
better than the one obtained in previous studies of these models.

In summary, we conclude that in the LRSM it is possible to obtain from
the experimental results a value for $N_\nu$ different from 3 (not
necessarily an integer number). In particular for the left-right
symmetric model with Dirac neutrinos, $(N_\nu)_{LRSM}$ is in the
neighborhood of three. However, if new precision experiments find
small deviations from three, this model may explain very well these
deviations with a small value of $\phi$. We have shown that new
data of $R_{exp}=\frac{\Gamma_{inv}}{\Gamma_{l\bar l}}$ can
considerably shrink the allowed region of $(N_{\nu})_{LRSM}$. In
the limit of $\phi=0$, our bounds takes the value previously
reported in the literature \cite{Data06,Abbaneo,M.Carena}.


\vspace{2cm}

\begin{center}
{\bf Acknowledgments}
\end{center}

We would like to thank O. G. Miranda for useful discussions. This work was supported
by CONACyT and SNI (M\'exico).

\newpage

\begin{figure}[t]
\centerline{\scalebox{0.85}{\includegraphics{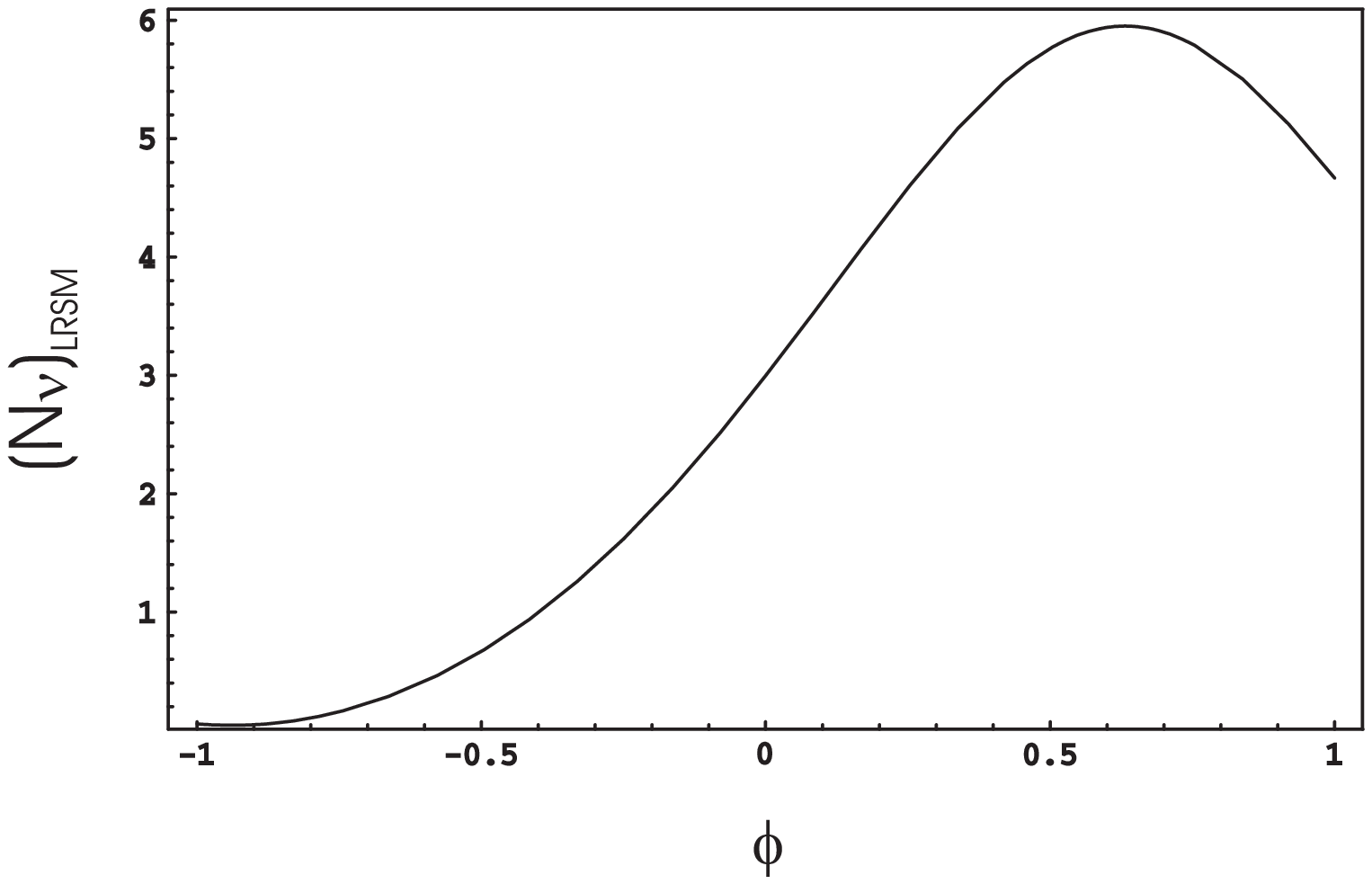}}}
\caption{ \label{fig:gamma} $(N_\nu)_{LRSM}$ as a function of the
mixing angle $\phi$.}
\end{figure}

\begin{figure}[t]
\centerline{\scalebox{0.85}{\includegraphics{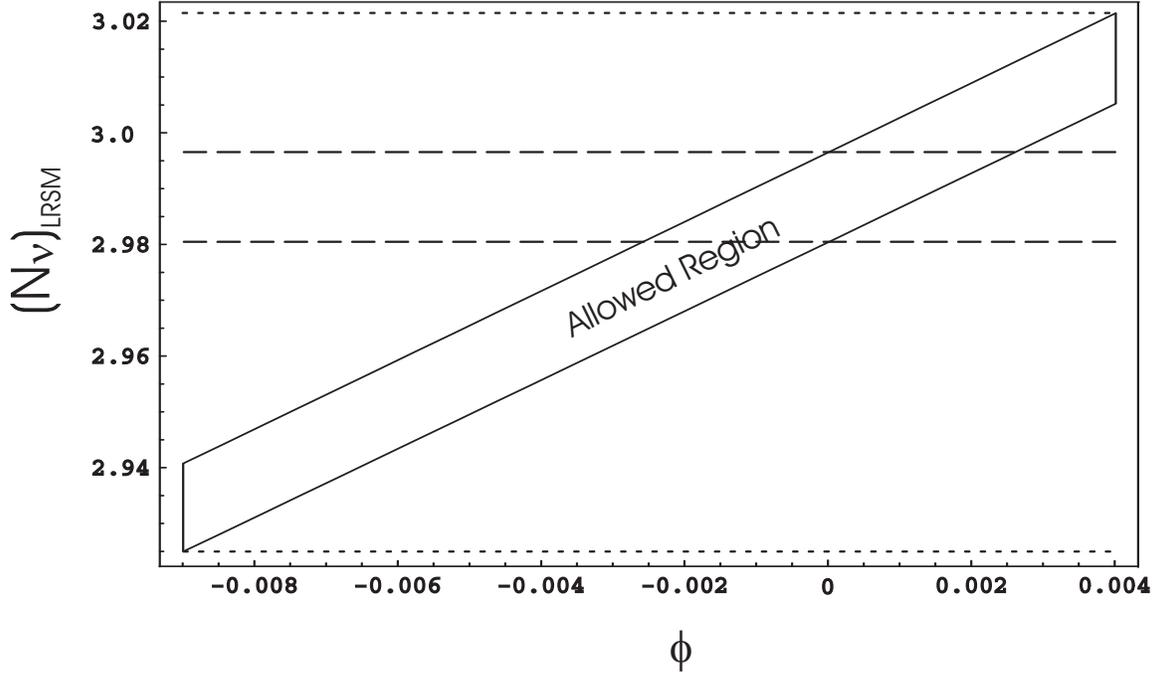}}}
\caption{ \label{fig:gamma} Allowed region for $(N_\nu)_{LRSM}$ as
a function of the mixing angle $\phi$ with the experimental value
$R^{LEP}_{exp}$. The dashed line shows the SM allowed region for
$N_\nu$ at $90\%$ C.L., while the dotted line shows the same
result for the LRSM.}
\end{figure}

\begin{figure}[t]
\centerline{\scalebox{0.85}{\includegraphics{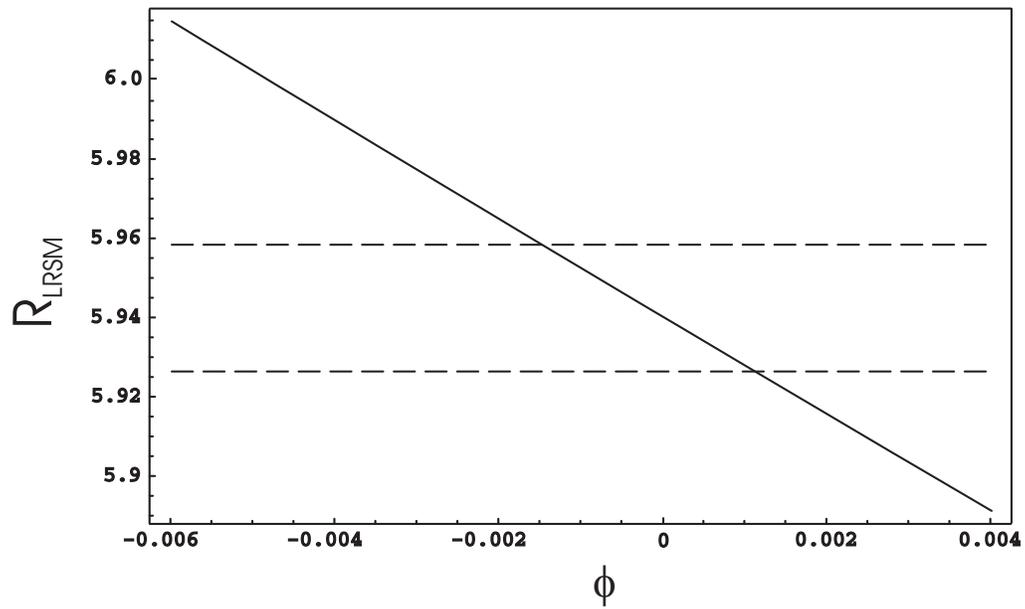}}}
\caption{ \label{fig:gamma} The curve shows the shape for
$R_{LRSM}$ as a function of the mixing angle $\phi$. The dashed
line shows the experimental region for $R^{LEP}_{exp}$ at $90\%$
C.L..}
\end{figure}

\end{document}